\documentclass[aps,floats,floatfix,showpacs,prd,superscriptaddress,onecolumn,nofootinbib]{revtex4}
\usepackage{graphicx, epsfig, amssymb} 
\usepackage{amsmath, amsfonts}
\usepackage{bm} 
\usepackage{color}
\usepackage{mathtools}
\usepackage{enumerate}
\usepackage[T1]{fontenc}
\usepackage{float}
\usepackage{array}
\usepackage{multirow}
\usepackage[multiple]{footmisc}

\makeatletter

\@ifundefined{textcolor}{}
{%
 \definecolor{BLACK}{gray}{0}
 \definecolor{WHITE}{gray}{1}
 \definecolor{RED}{rgb}{1,0,0}
 \definecolor{GREEN}{rgb}{0,1,0}
 \definecolor{BLUE}{rgb}{0,0,1}
 \definecolor{CYAN}{cmyk}{1,0,0,0}
 \definecolor{MAGENTA}{cmyk}{0,1,0,0}
 \definecolor{YELLOW}{cmyk}{0,0,1,0}
}

\makeatother

\newcommand{\be}{\begin{equation}}
\newcommand{\ee}{\end{equation}}                  
\newcommand{\bea}{\begin{eqnarray}}
\newcommand{\eea}{\end{eqnarray}}

\begin{document}


\title{Joule-Thomson expansion of AdS black holes in quasitopological electromagnetism}

\author{Jos\'e Barrientos}
\email{barrientos@math.cas.cz}
\affiliation{Institute of Mathematics of the Czech Academy of Sciences, \v{Z}itn\'a 25, 11567 Praha 1, Czech Republic}
\affiliation{Departamento de Ense\~nanza de las Ciencias B\'asicas, Universidad Cat\'olica del Norte, Larrondo 1281, Coquimbo, Chile}
\author{Jos\'e Mena}
\email{jomena@udec.cl}
\affiliation{Departamento de F\'isica, Universidad de Concepci\'on, Casilla 160-C, Concepci\'on, Chile}

\noindent 
\begin{abstract}
We study the Joule-Thomson expansion in Einstein-Maxwell theory supplemented with the so-called quasitopological electromagnetism, this in the extended phase space thermodynamic approach. We compute the Joule-Thomson coefficient and depict all relevant inversion and isenthalpic curves in the temperature-pressure plane, determining in this manner the corresponding cooling and heating regions. In contrast with previous related works we show the existence of three branches for the inversion curves which depends on suitable selections of the parameter space of the theory, thus departing from the usual van der Waals behavior which exhibits up to two branches.
\end{abstract}


\maketitle


\section{Introduction}
Black hole thermodynamics have been extensively studied since the pioneering works of Bekenstein and Hawking \cite{Bekenstein:1972tm,Bekenstein:1973ur,Bekenstein:1974ax,Bardeen:1973gs,Hawking:1974rv} and it has been shown to possess fundamental connections with classical thermodynamics, general relativity and quantum mechanics, providing a deep insight into the understanding of a quantum theory of gravity.
In recent decades, several works have been developed along these lines establishing a complete theory of black hole thermodynamics, opening the road for the study of black holes as thermodynamic systems exhibiting a wide range of phenomena from which phase transitions and black hole radiation are specially highlighted \cite{tHooft:1984kcu,Cardy:1986ie,Zhang:2005xt}. 
The study of black hole thermodynamics in the presence of a negative cosmological constant is particularly appealing \cite{Witten:1998qj} since its rich phase structure and its dual description based on the AdS/CFT correspondence \cite{Maldacena:1997re}. 
Hawking and Page \cite{Hawking:1982dh} discovered the existence of thermodynamic phase transitions between Schwarzschild anti-de Sitter (AdS) black holes with large radii and thermal AdS space, which was demonstrated to be dual to a confinement/deconfinement phase transition in the boundary conformal field theory \cite{Witten:1998qj, Witten:1998zw}. In particular, charged static black holes with the AdS asymptotic obey an equation of state qualitatively similar to the van der Waals fluids, and exhibit first-order phase transitions between small and large black holes \cite{Chamblin:1999tk, Chamblin:1999hg, Rajagopal:2014ewa}. Several works have reported similar behavior for different charged AdS black hole configurations \cite{Salti:2021jpe, Roy:2021ucl, Balart:2021glm, Kubiznak:2016qmn, Caceres:2015vsa, Hennigar:2015esa}. 

When considering the possibility of a varying pressure \cite{Kubiznak:2012wp}, an identification that follows from a cosmological perspective since a negative cosmological constant induces a vacuum pressure, the mass $M$ of the black hole is interpreted as the enthalpy $H$ rather than the internal energy $U$ \cite{Kastor:2009wy, Dolan:2011xt, Kubiznak:2016qmn}.
Hence, the natural first law for black holes with cosmological constant includes besides the electromagnetic charges and rotation, a contribution arising from the nonzero energy of the cosmological constant contained in the volume of the black hole. The interpretation of the mass of black holes as enthalpy and the inclusion of the cosmological constant as a pressure have remarkable consequences for black hole thermodynamics, aligning them much closer with familiar phase spaces of chemical reactions. Indeed, in the extended phase space in which we consider the cosmological constant $\Lambda$ as pressure $P$,

\begin{equation}
P=-\frac{\Lambda}{8\pi},
\end{equation}
where its thermodynamic conjugate variable defines a thermodynamical volume
\begin{equation}
V=\left(\frac{\partial H}{\partial P}\right).
\end{equation}

In \cite{Kubiznak:2012wp} Kubiz\v{n}\'ak and Mann showed that phase transitions of charged AdS black holes are similar to those of van der Waals fluids and that these black hole shares $P -V$ diagram and critical behavior with  the van der Waals system. In recent years, the $P-V$ behavior of different charged AdS black holes has been extensively investigated \cite{Spallucci:2013osa,Gunasekaran:2012dq,Hendi:2012um,Chen:2013ce,Zhao:2013oza,Zou:2013owa,Zhang:2014jfa,Zou:2014mha,Frassino:2014pha,Hendi:2016vux}.

An interesting process in classical thermodynamics is the so-called Joule-Thomson (JT) expansion \cite{johnstonbook}. The Joule-Thomson effect is an isenthalpic process in which temperature change is produced when a gas is allowed to expand from a high pressure region to a low pressure region, this through a valve or porous plugs. 
As a result it is possible to obtain heating or cooling effects due to this process, where  both are controlled by an inversion point characterized by the so-called Joule-Thomson coefficient. This inversion point is defined as the point where the inversion curves intersect the isenthalpic curves in the $ T - P$ plane. The JT expansion for black holes was first studied by \"Okc\"u and Ayd\i{}ner in \cite{Okcu:2016tgt} for the case of Reissner-­Nordstr\"om AdS black holes finding similarities with the van der Waals fluid. They also presented the inversion curves that separate heating and cooling regions. The interest in studying the JT expansion for black holes is due to the fact that Hawking radiation can be interpreted as an adiabatic expansion even though there is no porous plug \cite{MahdavianYekta:2019dwf}.
These studies were then generalized to Kerr-AdS black holes \cite{Okcu:2017qgo}, AdS black holes with a global monopole \cite{RizwanCL:2018cyb}, to charged AdS black holes in $f(R)$ gravity \cite{Chabab:2018zix}, Lovelock gravity \cite{Mo:2018qkt}, Gauss-Bonnet gravity \cite{Lan:2018nnp}, Einstein-Maxwell-axion and massive gravity \cite{Cisterna:2018jqg}. All the works show that the inversion curves in the $T - P$ plane have positive slope only while for van der Waals system have both positive and negative slopes. Some works also investigated this phenomenon in nonlinear electrodynamics systems \cite{Li:2019jcd,Nam:2019zyk,Bi:2020vcg,Feng:2020swq,Kruglov:2022lnc} reproducing similar behavior. Due to these results, it would be expected that all AdS spacetimes present a Joule-Thomson related phenomena. Nonetheless  in AdS spacetimes \cite{Chougule:2018cny, Pourhassan:2017qxi} there is no van der Waals like behavior, which can lead to the absence of Joule-Thomson expansion.

The aim of this paper is to present the study of the Joule-Thomson effect in quasitopological electromagnetism \cite{Liu:2019rib}, a theory that corresponds to a higher-order extension of Maxwell electromagnetism constructed with the bilinear invariant $F^2=F_{\mu\nu}F^{\mu\nu}$. In \cite{Liu:2019rib} the authors presented the corresponding charged AdS black hole. Although the new terms have a nonvanishing contribution to the field equations, they do not affect the purely electric or magnetic Reissner-Nordstr\"om solutions. However,  they affect the dyonic black hole solution,  endowing the spacetime with four black hole horizons for certain parameter regions. It was also found that there  exist three photon spheres, with  one of them being stable. For black hole solutions, the matter sector satisfies the dominant energy condition, and also the null and weak energy conditions, but may violate the strong energy condition. Another remarkable characteristic is that the energy-momentum tensor of the quasitopological contribution is of the form of isotropic perfect fluid, with isotropic pressure being accurately the opposite of the energy density. In this manner the quasitopological term provides a candidate for dark energy. In Ref. \cite{Li:2022vcd} the authors presented the thermodynamics phase transitions of the solutions. 

This work is organized as follows. Section \ref{section bhthermo} is devoted to presenting the black hole solutions of four-dimensional Einstein gravity minimally coupled to Maxwell and quasitopological electromagnetism and to analyzing their thermodynamical properties. In Sec. \ref{sectionJT}  the Joule-Thomson expansion is investigated for black holes in quasitopological electromagnetism. Finally, we conclude our results and discuss possible future directions in Sec. \ref{sectionconclusion}.

\section{Black hole thermodynamics}\label{section bhthermo}
The four-dimensional action of the minimally coupled Einstein-Maxwell-quasitopological electromagnetism is
\begin{equation}
S=\int dx^4\sqrt{-g}\left(R-2\Lambda-\alpha_1 F^2-\alpha_2\left(\left(F^2\right)^2-2F^{(4)}\right)\right),
\end{equation} 
where  
\begin{equation}
\begin{aligned}
F^2&=F_{\mu\nu}F^{\mu\nu},\\
F^{(4)}&=F^\mu_{\,\,\,\nu}F^\nu_{\,\,\,\rho}F^\rho_{\,\,\,\lambda}F^\lambda_{\,\,\,\mu},
\end{aligned}
\end{equation}
with $F_{\mu\nu}$ the strength of the electromagnetic field. The Einstein field equations and Maxwell equations are  respectively given by
\begin{equation}
\begin{aligned}
R_{\mu\nu}-\frac{1}{2}g_{\mu\nu}R+\Lambda g_{\mu\nu}&=T_{\mu\nu},\\
\nabla_\mu \tilde{F}^{\mu\nu}&=0,
\end{aligned}
\end{equation}
where the energy-momentum tensor of the system is
\begin{equation}
\begin{aligned}
T_{\mu\nu}&=\alpha_1 T^{(1)}_{\mu\nu}+\alpha_2 T^{(2)}_{\mu\nu},\\
T^{(1)}_{\mu\nu}&=2F_{\mu\lambda}F_{\nu}^{\,\,\,\lambda}-\frac{1}{2}g_{\mu\nu}F^2,\\
T^{(2)}_{\mu\nu}&=4F^2 F_{\mu\lambda}F_\nu^{\,\,\,\lambda}-8F_{\mu\lambda}F^\lambda_{\,\,\,\rho}F^\rho_{\,\,\,\sigma}F^\sigma_{\,\,\,\nu}-\frac{1}{2}g_{\mu\nu}\left(\left(F^2\right)^2-2F^{(4)}\right),
\end{aligned}
\end{equation}
and
\begin{equation}
\tilde{F}^{\mu\nu}=4\alpha_1 F^{\mu\nu}+8\alpha_2\left(F^2F^{\mu\nu}-2F^{\mu\lambda}F^\rho_{\,\,\,\lambda}F_\rho^{\,\,\,\nu}\right).
\end{equation}
The parameters $\alpha_1,\,\alpha_2$ are coupling constants where $\alpha_1$ is dimensionless and  $\alpha_2$ is of dimension $(\text{length})^2$. The  related Hamiltonian  is non-negative provided that $\alpha_{1,2}>0$ \cite{Liu:2019rib}. Therefore, from here on we assume that $\alpha_1,\,\alpha_2$ are positive. 
The equations of motion admit a spherically symmetric dyonic black hole solution:
\begin{equation}
ds^2=-f(r)dt^2+\frac{dr^2}{f(r)}+r^2(d\theta^2+\sin^2\theta d\phi^2),\quad F=-a^\prime (r) dt\wedge dr+\alpha_1 Q_m \sin\theta d\phi \wedge d\theta, 
\end{equation}
where the metric function is given by
\begin{equation}
f(r)=-\frac{1}{3}\Lambda r^2+1-\frac{2M}{r}+\frac{\alpha_1^3 Q_m^2}{r^2}+\frac{Q_e^2}{\alpha_1 r^2} {_2}F_1[\frac{1}{4},1,\frac{5}{4},-\frac{4\alpha_1\alpha_2Q_m^2}{r^4}],
\end{equation}
where $M$ is the mass of the black hole, $Q_e,\, Q_m$ the electric and magnetic charges. The electric potential satisfies
\begin{equation}
a^\prime (r)=-\frac{Q_e r^2}{\alpha_1\left(r^4+4\alpha_1\alpha_2 Q_m^2\right)}.
\end{equation}
It is now possible to write the thermodynamical quantities related to the black hole \cite{Li:2022vcd}. Considering $r_+$ as the largest root of $f(r_+)=0$ and the negative cosmological constant as positive pressure $P=-\frac{\Lambda}{8\pi}$ then the temperature and the entropy are
\begin{equation}\label{eqT}
\begin{aligned}
T&=\frac{1}{4\pi r_+}+2Pr_+-\frac{\alpha_1^3Q_m^2}{4\pi r_+^3}-\frac{Q_e^2 r_+}{4\pi\alpha_1\left(r_+^4+4\alpha_1\alpha_2 Q_m^2\right)},\\
S&=\pi r_+^2.
\end{aligned}
\end{equation}
Black hole mass, electric and magnetic potentials are given by
\begin{equation}\label{eqmasa}
\begin{aligned}
M&=\frac{3\alpha_1 r_+^2+8\pi \alpha_1 P r_+^4+3\alpha_1^4 Q_m^2+3Q_e^2 {}_2F_1[\frac{1}{4},1,\frac{5}{4},-\frac{4\alpha_1\alpha_2 Q_m^2}{r_+^4}]}{6\alpha_1 r_+},\\
\Phi_e&=\frac{Q_e {}_2F_1[\frac{1}{4},1,\frac{5}{4},-\frac{4\alpha_1\alpha_2Q_m^2}{r_+^4}]}{\alpha_1 r_+},\\
\Phi_m&=\frac{\alpha_1^3 Q_m}{r_+}+\frac{Q_e^2 r_+^3}{4\alpha_1 Q_m\left(r_+^4+4\alpha_1\alpha_2 Q_m^2\right)}-\frac{Q_e^2 {}_2F_1[\frac{1}{4},1,\frac{5}{4},-\frac{4\alpha_1\alpha_2Q_m^2}{r_+^4}]}{4\alpha_1 Q_m r_+}.
\end{aligned}
\end{equation}
It is then straightforward to verify the first law of thermodynamics:
\begin{equation}
dM=TdS+\Phi_edQ_e+\Phi_mdQ_m+VdP.
\end{equation}
The Smarr relation breaks because of the dimensionful of the coupling constant $\alpha_2$. To recover the latest relation we have to  treat $\alpha_2$ as a thermodynamical variable and therefore we obtain the first law of thermodynamics as follows:
\begin{equation}
dM=TdS+\Phi_edQ_e+\Phi_mdQ_m+VdP+\Phi_{\alpha_2}d\alpha_2
\end{equation}
and the corresponding Smarr relation
\begin{equation}
M=2TS-2PV+\Phi_eQ_e+\Phi_mQ_m+2\alpha_2\Phi_{\alpha_2},
\end{equation}
where $\Phi_{\alpha_2}$ is the conjugate quantity to $\alpha_2$,
\begin{equation}
\Phi_{\alpha_2}=\frac{Q_e^2r_+^3}{8\alpha_1\alpha_2(r_+^4+4\alpha_1\alpha_2Q_m^2)}-\frac{Q_e^2 {}_2F_1[\frac{1}{4},1,\frac{5}{4},-\frac{4\alpha_1\alpha_2Q_m^2}{r_+^4}]}{8\alpha_1\alpha_2 r_+}.
\end{equation}
\section{Joule-Thomson expansion}\label{sectionJT}
\begin{figure}[H]
 \centering
 \includegraphics[scale=0.9]{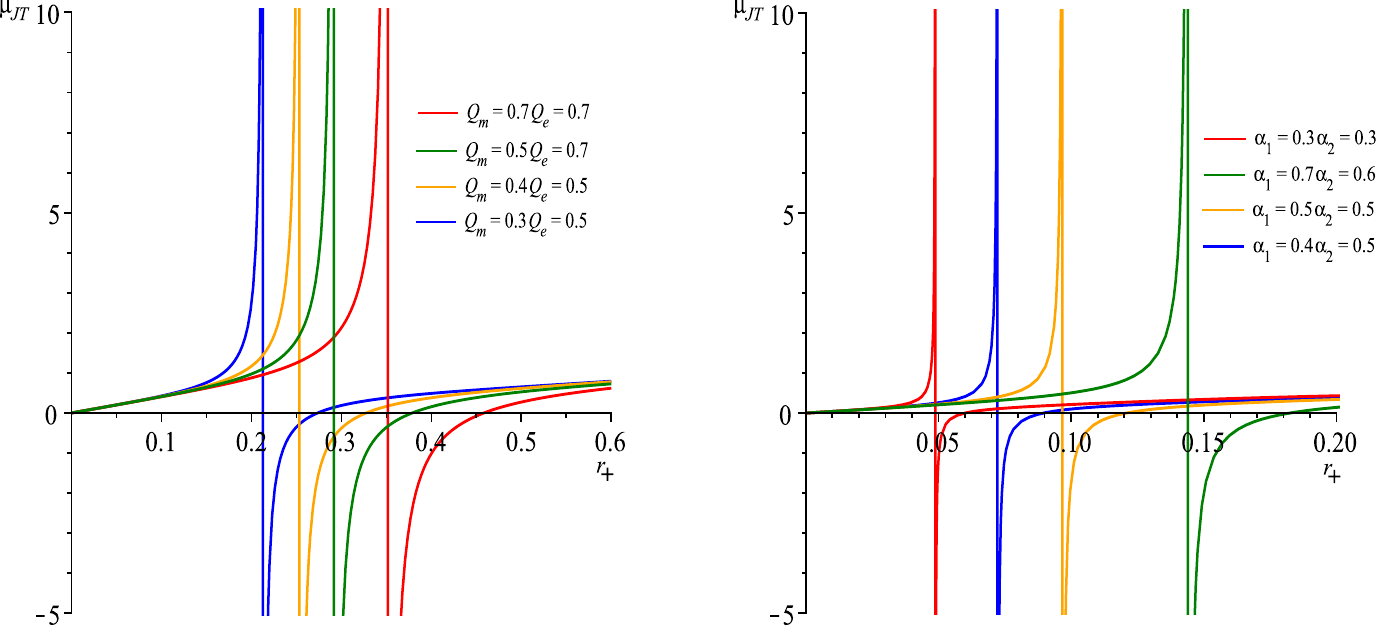}
 \caption{The plot of Joule-Thomson coefficient $\mu_{JT}$ with respect to $r_+$ at $P=1$. On the left,   fixed values $\alpha_1=1,\, \alpha_2=0.2$ for different values of the charges $Q_e,\, Q_m$ were considered. On the right, fixed values $Q_e=0.4, Q_m=0.3$ for different values of $\alpha_1,\, \alpha_2$.}
 \label{figuramjtvsr}
\end{figure}

\begin{figure}[H]
 \centering
 \includegraphics[scale=0.9]{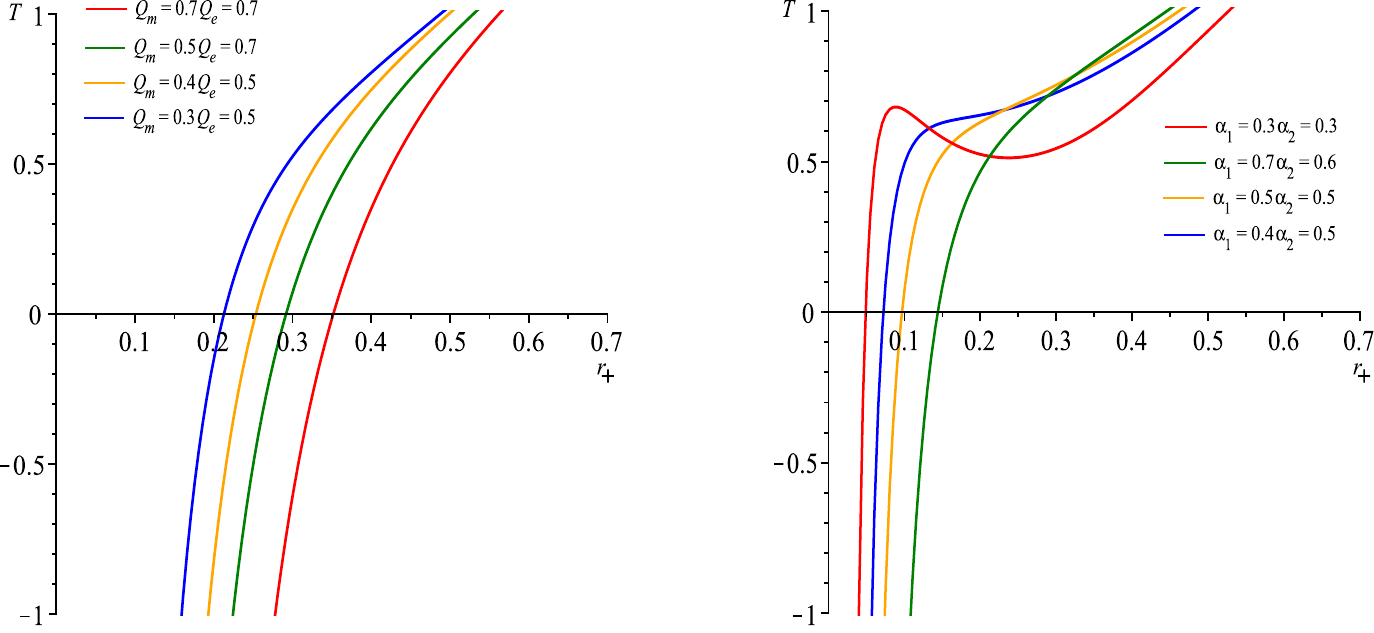}
  \caption{The plot of temperature $T$ with respect to $r_+$ at $P=1$. On the left,   fixed values $\alpha_1=1,\, \alpha_2=0.2$ for different values of the charges $Q_e,\, Q_m$ were considered. On the right, fixed values $Q_e=0.4, Q_m=0.3$ for different values of $\alpha_1,\, \alpha_2$.}
  \label{figuratvsr}
\end{figure}

In this section we study the Joule-Thomson expansion for the dyonic black hole into the quasitopological electromagnetism exposed in the previous section. The Joule-Thomson effect occurs when in an isenthalpic process, the temperature changes as  the gas expands from the  high pressure section to the low pressure through porous plugs \cite{Okcu:2016tgt}. The change of temperature with respect to pressure can be described by the Joule-Thomson coefficient defined as \cite{reif}
\begin{equation}\label{mujt}
\mu_{JT}=\left(\frac{\partial T}{\partial P}\right)_H=\frac{1}{C_P}\left[T\left(\frac{\partial V}{\partial T}\right)_P-V\right],
\end{equation}
where $C_P=T\left(\partial S/ \partial T\right)_P$ is the heat capacity at constant pressure. The sign of $\mu_{JT}$ determines whether heating or cooling will occur. In the JT expansion the change of pressure is negative but the change of temperature can be positive or negative. For $\mu_{JT}>0$, one has a cooling region in the $T-P$ plane whereas $\mu_{JT}<0$ determines the heating region in the $T-P$ plane. Replacing the thermodynamical quantities into Eq. \eqref{mujt} one finds
\begin{small}
\begin{equation}
\mu_{JT}=\frac{4r_+\left[8\pi\alpha_1 P(r_+^4+4\alpha_1\alpha_2 Q_m^2)^2r_+^4+8\alpha_1^2\alpha_2Q_m^2(2r_+^2-3\alpha_1^3Q_m^2)(r_+^4+2\alpha_1\alpha_2 Q_m^2)-3(Q_e^2+\alpha_1^4 Q_m^2)r_+^8+2\alpha_1(r_+^6-2\alpha_2Q_e^2Q_m^2)r_+^4\right]}{3(r_+^4+4\alpha_1\alpha_2 Q_m^2)\left[8\pi\alpha_1 P(r_+^4+4\alpha_1\alpha_2 Q_m^2)r_+^4- \alpha_1^2 Q_m^2 (\alpha_1^2 r_+^4-4\alpha_2r_+^2+4\alpha_1^3\alpha_2Q_m^2)+\alpha_1r_+^6-Q_e^2 r_+^4\right]}.
\end{equation}
\end{small}
The Joule-Thomson coefficient $\mu_{JT}$ versus the horizon $r_+$ is shown in Fig. \ref{figuramjtvsr} for different values of $Q_e,\, Q_m$ ($\alpha_1,\, \alpha_2$) fixing $\alpha_1,\, \alpha_2$ ($Q_e,\, Q_m$) at $P=1$. In both cases exist a divergent point and a zero point. When the event horizon $r_+$ is  large enough, the JT coefficient is positive and for decreasing $r_+$, $\mu_{JT}$ gradually decreases to zero and goes to negative. There is a $r_+$ for each case in which the coefficient $\mu_{JT}$ diverges. This point is in accordance with Fig. \ref{figuratvsr} where one can observe that corresponds to the points where $T=0$. 
By setting $\mu_{JT}=0$ one can define the inversion pressure $P_i$, the particular point in the gradient of pressure of the black hole for which the system changes from cooling (heating) to heating (cooling), and then replacing into Eq. \eqref{eqT} for the corresponding $T_i$ one obtains the following parametric equation for the inversion curves:
\begin{equation}\label{eqinversioncurves}
\begin{aligned}
T_{i}&=\frac{(2Q_e^2-\alpha_1r_+^2+2\alpha_1^4Q_m^2)r_+^8-8\alpha_1^2\alpha_2 Q_m^2(r_+^2-2\alpha_1^3Q_m^2)(r_+^4+2\alpha_1\alpha_2 Q_m^2)}{4\pi\alpha_1(r_+^4+4\alpha_1\alpha_2Q_m^2)^2r_+^3},\\ 
P_{i}&=\frac{3(Q_e^2+\alpha_1^4Q_m^2)r_+^8-8\alpha_1^2\alpha_2 Q_m^2(2r_+^2-3\alpha_1^3Q_m^2)(r_+^4+2\alpha_1\alpha_2 Q_m^2)-2\alpha_1(r_+^6-2\alpha_2Q_e^2Q_m^2)r_+^4}{8\pi\alpha_1(r_+^4+4\alpha_1\alpha_2Q_m^2)^2r_+^4}. 
\end{aligned}
\end{equation}

\begin{figure}[H]
 \centering
 \includegraphics[scale=0.9]{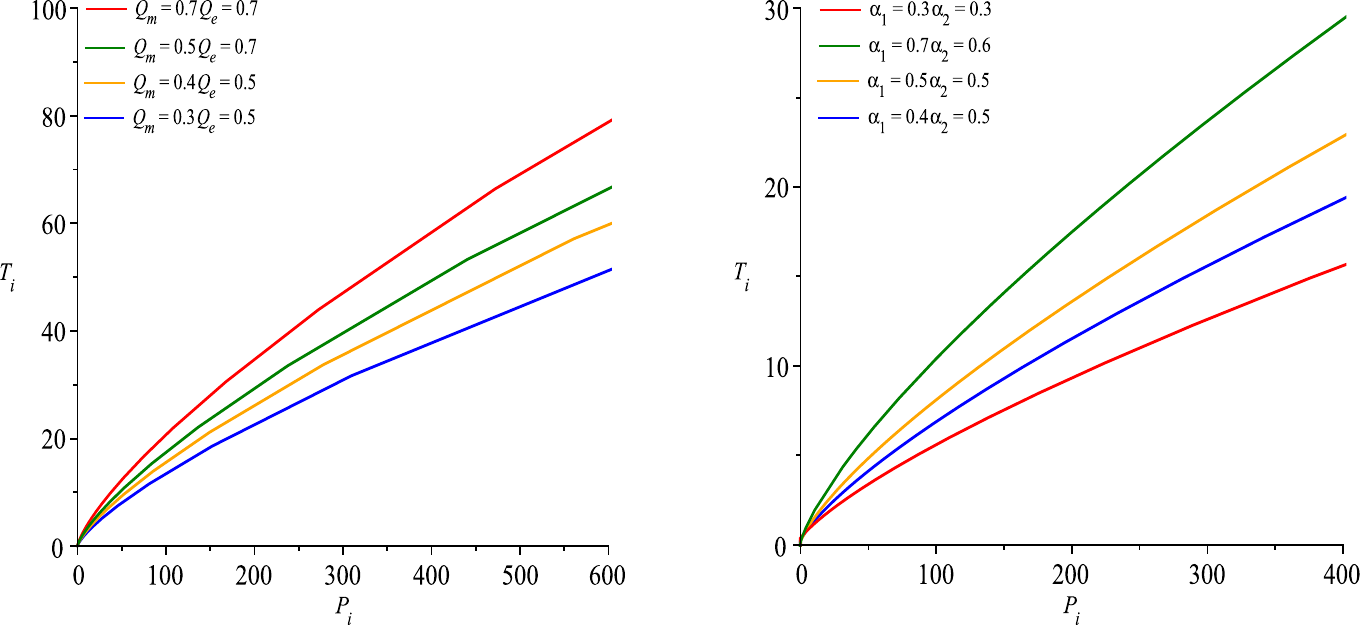}
 \caption{Inversion curves $T_i-P_i$. On the left,   fixed values $\alpha_1=1,\, \alpha_2=0.2$ for different values of the charges $Q_e,\, Q_m$. On the right, fixed values $Q_e=0.4, Q_m=0.3$ for different values of $\alpha_1,\, \alpha_2$.}
 \label{figuramtvsp}
\end{figure}

\begin{figure}[H]
 \centering
 \includegraphics[scale=0.9]{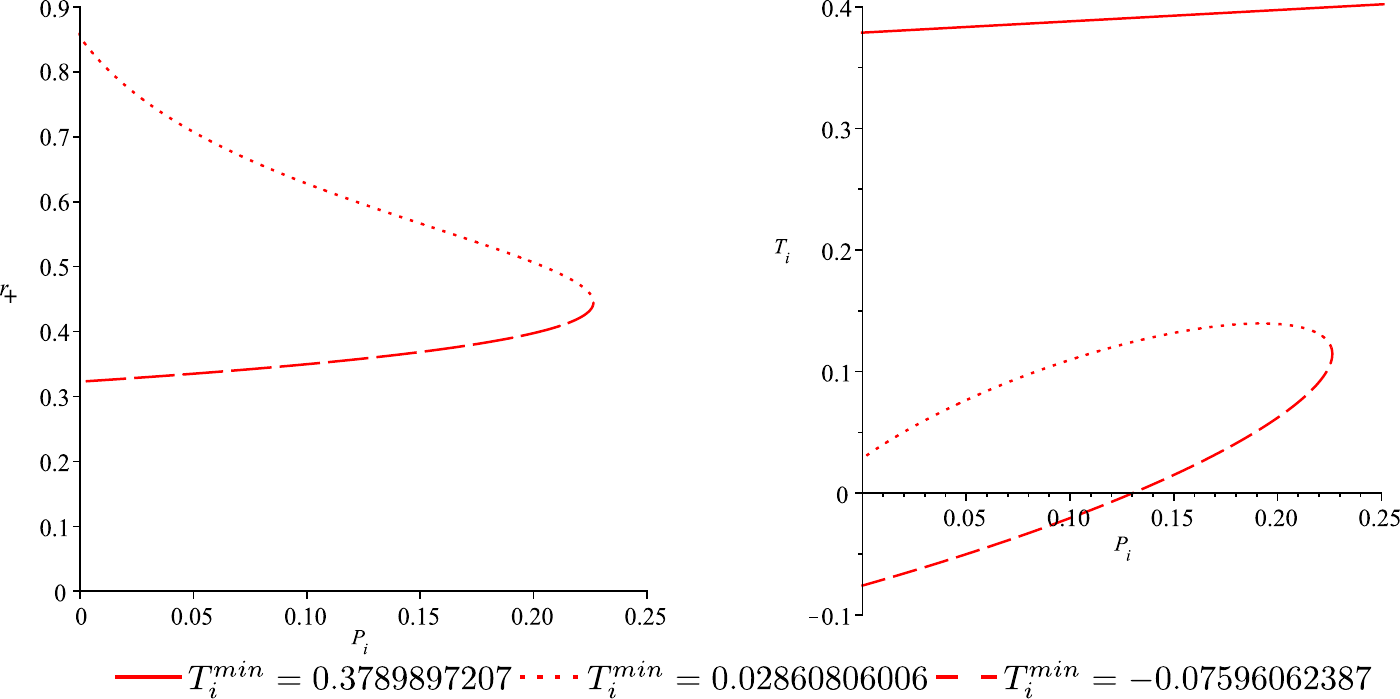}
 \caption{Left: positive $r_+$ versus inversion pressure for the lower branches. Right: inversion curve $T_i-P_i$. The  values are $Q_e=0.4,\, Q_m=0.3,\, \alpha_1=0.3,\,\alpha_2=0.3$.}
 \label{3branches}
\end{figure}

Inversion curves for different values of $Q_e,\,Q_m$ and $\alpha_1,\,\alpha_2$ are illustrated in Fig. \ref{figuramtvsp}. The inversion temperature  seems to  increase monotonically with the inversion pressure which means  the inversion curves are not closed and there is only one inversion curve in contrast to van der Waals fluids \cite{Okcu:2016tgt}. However, for some particular choice  of the parameters,  $Q_e=0.4,\, Q_m=0.3,\, \alpha_1=0.3,\,\alpha_2=0.3$, there exist three minimum values of the inversion temperature $T_i^{\text{min}}$ corresponding to the zero inversion pressure $P_i=0$. This case is shown in Fig. \ref{3branches} where the solid line curve represents the red curve  on    Fig. \ref{figuramtvsp} (right) which corresponds to $T_i^{\text{min}}=0.3789897207$. Figure \ref{3branches} (right) also shows the other two branches, the dotted line and dashed line correspond to $T_i^{\text{min}}=0.02860806006$ and $T_i^{\text{min}}=-0.07596062387$ respectively. On the left we show that the related horizons that satisfy $\mu_{JT}=0$ are positive and one can verify that for the three cases the corresponding mass of the black hole is positive. This behavior is novelty and  significantly differs from the studied black holes \cite{Okcu:2016tgt, Okcu:2017qgo,RizwanCL:2018cyb,Chabab:2018zix,Mo:2018qkt,Lan:2018nnp,Cisterna:2018jqg,Li:2019jcd,Nam:2019zyk,Bi:2020vcg,Feng:2020swq,Kruglov:2022lnc} and even from the van der Waals case. In k-essence models it was reported to exist two branches, but having negative mass of the black hole, making the Joule-Thomson expansion breaks down \cite{Cisterna:2018jqg}.
The cooling region and the heating region are located above and below these curves, respectively. Considering that the Joule-Thomson expansion is an isenthalpic process one can plot  the isenthalpic curves in the $T-P$ plane. Taking into account Eqs. \eqref{eqT} and \eqref{eqmasa}, the parametric equation for isenthalpic  process is given by
\begin{equation}\label{eqisenthalpic}
\begin{aligned}
T&=-\frac{1}{2\pi r_+}+\frac{3M}{2\pi r_+^2}-\frac{\alpha_1^3Q_m^2}{\pi r_+^3}-\frac{3Q_e^2 \Phi[-\frac{4\alpha_1\alpha_2Q_m^2}{r_+^4},1,\frac{1}{4}]}{16\pi\alpha_1 r_+^3}-\frac{Q_e^2 r_+}{4\pi\alpha_1(r_+^4+4\alpha_1\alpha_2Q_m^2)},\\
P&=-\frac{3}{8\pi r_+^2}+\frac{3M}{4\pi r_+^3}-\frac{3\alpha_1^3Q_m^2}{8\pi r_+^4}-\frac{3Q_e^2\Phi[-\frac{4\alpha_1\alpha_2Q_m^2}{r_+^4},1,\frac{1}{4}]}{32\pi\alpha_1r_+^4},
\end{aligned}
\end{equation}
where $\Phi[z,s,\alpha]$ is the Lerch transcendent function.

\begin{figure}[H]
 \centering
 \includegraphics[scale=0.9]{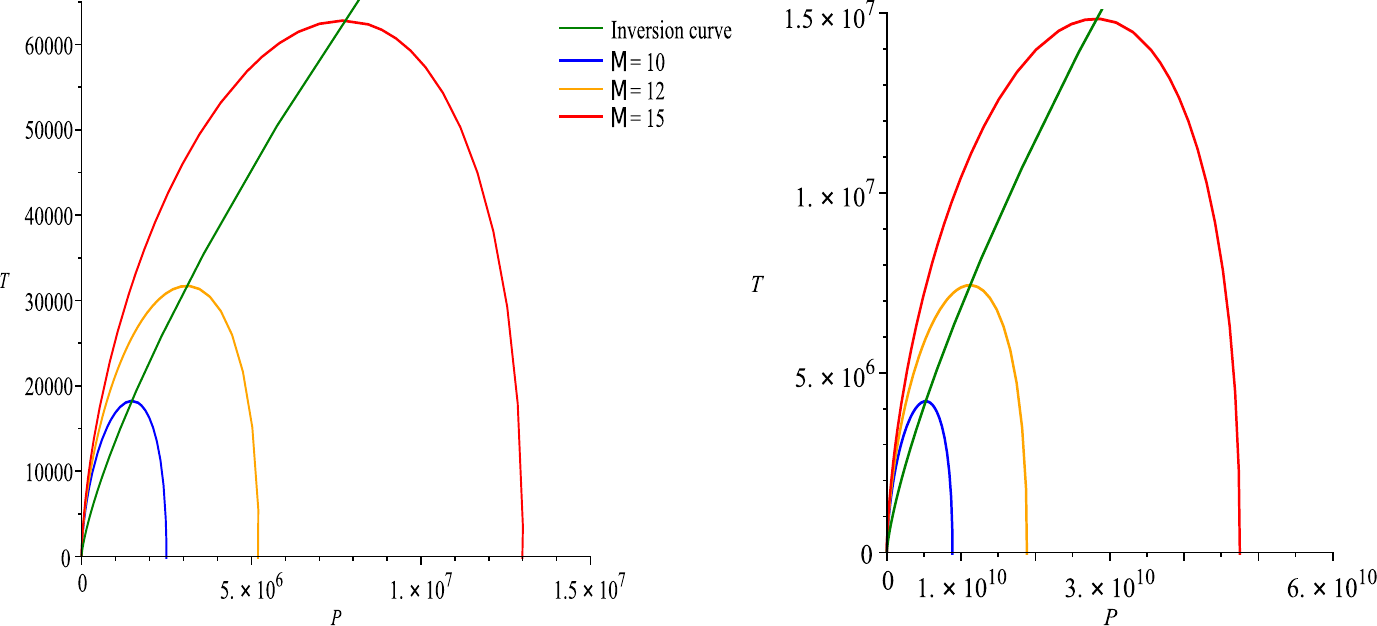}
 \caption{Isenthalpic curves for different values of $M$. Its set $Q_e=0.5,\, Q_m=0.3,\,\alpha_1=1,\,\alpha_2=0.2$ on the left. On the right $Q_e=0.4,\, Q_m=0.3,\,\alpha_1=0.4,\,\alpha_2=0.5$.}
 \label{figuraisenthalpic}
\end{figure}

By considering the  mass of the black hole equal to the enthalpy in the extended phase space \cite{Kastor:2009wy}, the isenthalpic curves for different values of the mass are plotted in Fig. \ref{figuraisenthalpic}. The inversion curve intersects the maximum point of the isenthalpic curves and divides them into two parts,  cooling and heating. For $P<P_i$, the slope of the isenthalpic curve is positive and then a cooling happens in the expansion. For $P>P_i$, the sign of the slope changes under the inversion curves and heating occurs.
\begin{figure}[H]
 \centering
 \includegraphics[scale=0.9]{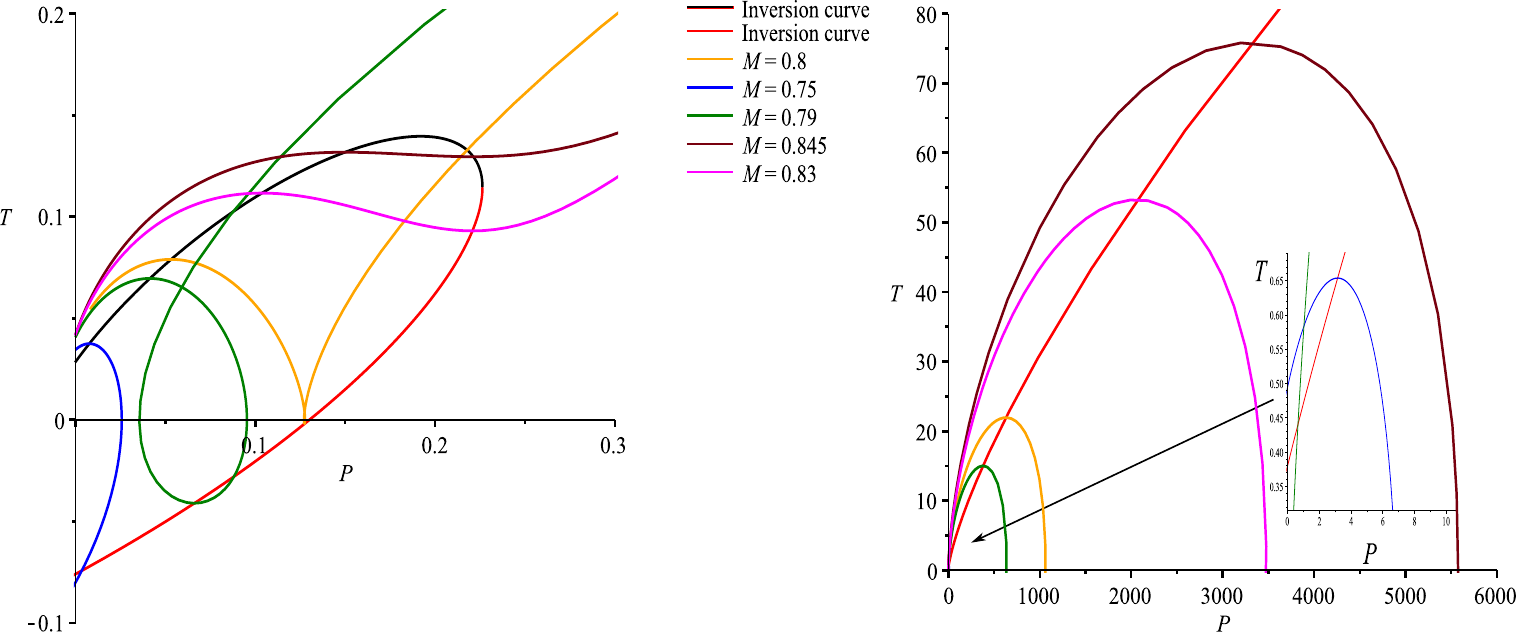}
 \caption{Isenthalpic curves for $Q_e=0.4,\, Q_m=0.3,\, \alpha_1=0.3,\,\alpha_2=0.3$ for different masses $M$ and the corresponding inversion curve.}
 \label{characteristic}
\end{figure}
The characteristic case of  three branches that corresponds to $Q_e=0.4,\, Q_m=0.3,\, \alpha_1=0.3,\,\alpha_2=0.3$ is shown in Fig. \ref{characteristic} where the red and black lines are the inversion curves. From the blue line to the brown one the masses of the black hole are $M=0.75,\,0.79,\,0.80,\,0.83$ and $0.845$, and one can check that the related horizons are all positive. Some of the isenthalpic curves (blue, green and yellow) cross the lower branch, but being the minimal inversion temperature negative. The other two  isenthalpic curves (magenta and brown) possess a maximum and a minimum, behavior that also differs with respect to the known literature \cite{Okcu:2016tgt, Okcu:2017qgo,RizwanCL:2018cyb,Chabab:2018zix,Mo:2018qkt,Lan:2018nnp,Cisterna:2018jqg,Li:2019jcd,Nam:2019zyk,Bi:2020vcg,Feng:2020swq,Kruglov:2022lnc} and the van der Waals fluids. These two points where $\mu_{JT}=0$ means that as usual the cooling process where $\mu_{JT}>0$ occurs above the inversion curve while the warming process where $\mu_{JT}<0$ occurs in the right side of the inversion curve.
\section{Conclusions and further developments}\label{sectionconclusion}
By considering the cosmological constant as a thermodynamical quantity we have analyzed the Joule-Thomson expansion; this means the expansion of gas from a higher pressure section to a lower one by maintaining the enthalpy (identified as the mass of the black hole) of the process constant, this in the context of charged AdS black holes in quasitopological electromagnetism. We have studied the Joule-Thomson coefficient $\mu_{JT}$ to determine the cooling and heating regions. We computed the inversion curves in the $T_i - P_i$ plane and the corresponding isenthalpic curves. The inversion curve divides the $T - P$ plane into two regions. Above the inversion curve we obtained the cooling region while the region below the inversion curves corresponds to the warming one. 

We have found an interesting and new behavior of the process with respect to the inversion curves in previous works. For certain values of the parameters, the inversion curves possess three branches in contrast to what have been found in \cite{Okcu:2016tgt, Okcu:2017qgo,RizwanCL:2018cyb,Chabab:2018zix,Mo:2018qkt,Lan:2018nnp,Cisterna:2018jqg,Li:2019jcd,Nam:2019zyk,Bi:2020vcg,Feng:2020swq,Kruglov:2022lnc} differing from the case of van der Waals fluids where two branches exist.  The corresponding horizons to these three branches are all positive while one of the branches possesses negative inversion temperature, the latter regarded as nonphysical. Due to the existence of three branches for some values of the parameters the isenthalpic curves cross all inversion curves. 
We found a process that moves from a cooling region into a heating region and that then moves to the upper branch where it again faces a cooling region and passes once again to a heating phase (see magenta and brown curves in Fig. \ref{characteristic}). 

It would be interesting to explore the Joule-Thomson expansion in the extended quasitopological electromagnetism defined in  \cite{Cisterna:2020rkc} in which  a new field   is introduced, represented by a $p$-form allowing to construct homogeneous charged black strings with or without a cosmological constant in arbitrary dimensions \cite{Cisterna:2021ckn}. Given the similarities between both electromagnetic theories it is expected to find analogies in the behavior of inversion and isenthalpic curves as the existence of three branches. We postpone these studies for future work.
\\
\paragraph{\textbf{ACKNOWLEDGMENTS}} 
We would like to thank Adolfo Cisterna for useful discussions and enlightening comments. The work of J. B. is supported by the  ``Programme to support prospective
human resources - post Ph.D. candidates''  of the Czech Academy of Sciences, Project No. L100192101. J. M. is partially funded by Agencia Nacional de Investigaci\'on y Desarrollo ANID through FONDECYT Regular  1210500.

\end{document}